\documentclass[]{spie}  %>>> use for US letter paper
\usepackage{amsmath,amsfonts,amssymb}
\usepackage{graphicx}
\usepackage[colorlinks=true, allcolors=blue]{hyperref}
\usepackage{flafter} 
\title{Cross-Slot Metal-Mesh Bandpass Filters for Far-Infrared Astronomy}

\author[a]{Joanna Perido}
\author[b]{Kevin Denis}
\author[b]{Jason Glenn}
\author[b]{Nicholas F. Cothard}
\author[b]{Manuel Quijada}
\author[b]{Jessica Patel}
\author[b]{Edward Wollack}
\author[b]{Tilak Hewagama}
\author[b]{Shahid Aslam}
\author[c]{Peter K. Day}
\affil[a]{University of Colorado Boulder}
\affil[b]{NASA Goddard Space Flight Center}
\affil[c]{Jet Propulsion Laboratory}

\authorinfo{Further author information: (Send correspondence to Joanna Perido)\\Joanna Perido: E-mail: joanna.perido@colorado.edu\\  Jason Glenn: E-mail: jason.glenn@nasa.gov}

\begin{document} 
\maketitle

\begin{abstract}
The far-infrared (IR) region is rich with information needed to characterize interstellar dust and to investigate the cold outer planets of the solar system and their icy moons. The proposed sub-orbital observatory the Balloon Experiment for Galactic INfrared Science (BEGINS) will utilize cryogenic instruments to map spectral energy distributions (SEDs) of interstellar dust in the Cygnus molecular cloud complex. A future high priority flagship mission Uranus Orbiter and Probe carrying a net flux radiometer (NFR) will study the in situ heat flux of the icy giants atmosphere to 10 bar pressure. These instruments require far-IR filters to define the instrument spectral bandwidths. Our ultimate goal is to define the instrument bands of BEGINS and the NFR with linear-variable filters (LVFs) and discrete-variable filters (DVFs). The LVFs and DVFs will be made of metal mesh band-pass filters (MMBF) comprised of a 100 nm thick gold film with cross-shaped slots of varying sizes along a silicon (Si) substrate with cyclic olefin copolymer (COC) anti-reflection (AR) coatings. We present our progress towards LVFs and DVFs with simulated and measured transmission of a room temperature, non-AR coated, single-band 44~$\mu$m MMBF filter. We have successfully fabricated, measured, and modeled a non-AR coated, room temperature 44~$\mu$m MMBF. The transmission at room temperature and non-AR coated was measured to be 27\% with a resolving power of 11. When COC-AR coated on both sides the transmission is expected to increase to 69\% with a resolving power of 10. 
% Our goal is to define the instrument bands with continuously linear-variable filters (LVFs). Our LVFs are metal mesh band-pass filters (MMPBF) comprised of a 100 nm thick gold film with cross-shaped apertures of varying sizes along a silicon (Si) substrate with cyclic olefin copolymer (COC) anti-reflection (AR) coatings. Before fabricating an LVF, we first simulated and measured the transmission of a single 44~$\mu$m MMBF filter. We present our progress with simulated and measured transmission of a room temperature non-AR coated 44~$\mu$m MMBF filter. (Note: still need to include sentence on results)
% will be optimized for rapid spectral imaging with a spectral resolving power of R~7 from 25 to 65 μm and R of 3-6 from 65 to 250 μm to map IR SEDs in the Cygnus molecular cloud complex. While Thermal Imager for Europa Reconnaissance and Science is an instrument... Both these instruments require This will be achieved with continuously linear-variable filters (LVFs). Our LVFs are metal mesh band-passes (MMPB) comprised of a 100 nm thick gold film with cross-shaped apertures of varying sizes along a silicon (Si) substrate with an anti-reflection (AR) coating. We present our progress with simulated and measured transmission of room temperature non-AR coated 44 μm MMBF and prototype 17 mm long LVF from 24 to 36 μm.
 
\end{abstract}


\section{INTRODUCTION}
\label{sec:intro}  % \label{} allows reference to this section
Observations of far-IR spectra are a necessary means to characterize interstellar dust that plays an important role in understanding galaxy evolution and star formation. The proposed sub-orbital observatory the Balloon Experiment for Galactic INfrared Science (BEGINS)\cite{glenn2021begins} will utilize cryogenic instruments to map spectral energy distributions (SEDs) of interstellar dust in
the Cygnus molecular cloud complex with a resolving power ($R=\lambda/\Delta\lambda$) of $\sim$ 7 from 25-65~$\mu$m and of 3-6 from 70-250~$\mu$m. The far-IR region also provides astronomers with knowledge of planets' atmospheres and evolution. The future high priority flagship mission Uranus Orbiter and Probe will carry a net flux radiometer (NFR) to study the in situ heat flux of the icy giants atmosphere to 10 bar
pressure\cite{NAP26522,aslam2020advanced}. The NFR will measure Uranus's net radiation flux in seven spectral bands, spanning solar to infrared wavelengths (0.2 to 300~$\mu$m) \cite{aslam2020advanced}. Our ultimate goal is to define the instrument bands of BEGINS and the NFR with linear-variable filters (LVFs) and discrete-variable filters (DVFs). 
\par The LVFs and DVFs will be made of metal mesh band-pass filters (MMBF). MMBFs are frequency selective surfaces. Our MMBFs are comprised of metal thin-films with arrays of cross-slots (Fig. \ref{fig:mesh_geo}). MMBFs can be free-standing or deposited on a substrate. Our MMBFs are lithographically patterned on a silicon (Si) substrate. Research on MMBFs was first published by Ulrich in 1966 \cite{ulrich1967far} for far-IR wavelengths and have since been investigated for far-IR astronomy because they  are compact, present an easy fabrication process, their response can be tailored by changing the cross-slot parameters, and they can be cooled to liquid helium temperatures \cite{tomaselli1981far, porterfield1994resonant, moller1996cross, melo2012cross, merrell2012compact}.

\par Here, we present our work on a 44 ~$\mu$m MMBF. This is preliminary work towards LVFs and DVFs for both BEGINS and the NFR. The filter is made of 100 nm thick gold and will be supported by a silicon (Si) substrate. Gold was chosen as the thin-film because of its low resistivity which allows for high transmission. By making it 100 nm thick it is $\sim$5 times larger than the skin depth of gold at the desired bandpass frequency. We use high-resistivity floatzone Si wafers for low loss. In this paper, we present simulations to predict the transmission profile of the 44 ~$\mu$m MMBF with a specific cross-slot size. The simulations are compared to measurements of the fabricated filters. We also present and discuss a transmission-line model to model the transmittance of the MMBF. 

\par The paper is organized as follows: Section \ref{sec:params}, discusses the 44~$\mu$m MMBF design, Section \ref{sec:model}, discusses two model methods, simulations (\ref{sec:sims}) and a transmission-line model (\ref{sec:tlm}). Section \ref{sec:fab}, describes the filter fabrication process.  In Section \ref{sec:meas}, we discuss the filter measurements and compare the  measurements to the models. In the last section, Section \ref{sec:conc}, we add concluding remarks and discuss future work on filters for BEGINS and the NFR.

\section{Filter Cross-Slot Design}
\label{sec:params}
\begin{figure}[h!]
    \centering
    \includegraphics[width=.35\textwidth]{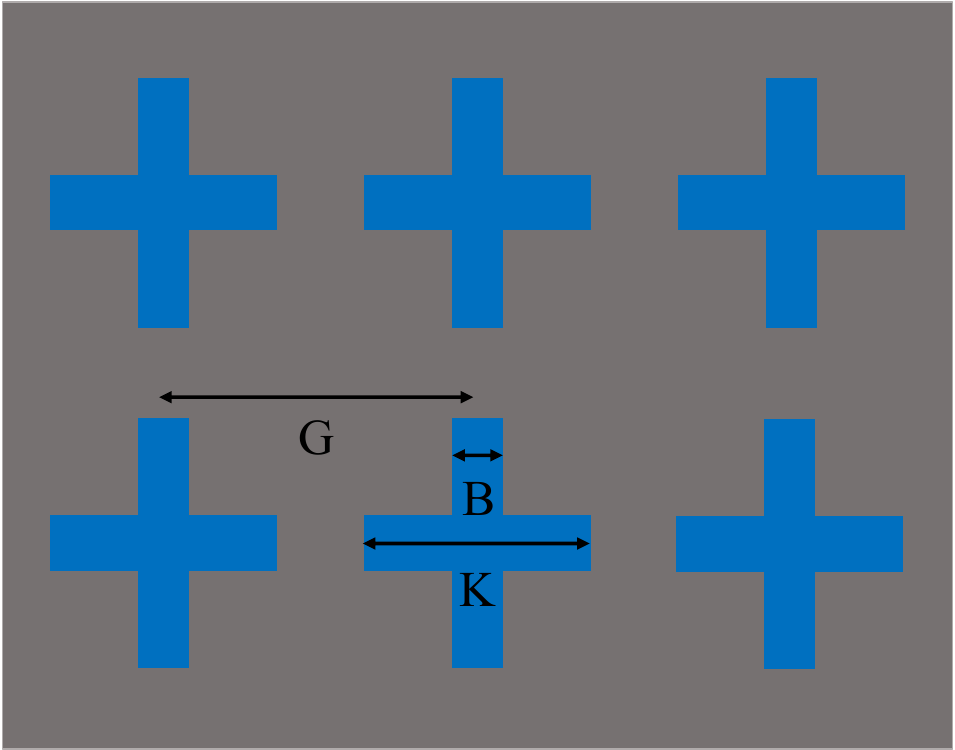}
    \caption{Illustration of metal-mesh cross-slot parameters. \textit{G} is the periodicity, \textit{K} is the cross-length and \textit{B} is the cross-width. Gray represents gold film. Blue represents bare Si substrate.}
    \label{fig:mesh_geo}
\end{figure}

The transmission profile of a MMBF is determined by its cross-slot parameters; the periodicity (G), the cross-length (K) and the cross-width (B)\cite{tomaselli1981far, porterfield1994resonant, moller1996cross, melo2012cross, merrell2012compact}, Fig. \ref{fig:mesh_geo}. The bandpass center peak scales with K and the bandwidth becomes small as the ratios of G/K and G/B increase\cite{porterfield1994resonant}.
The cross-slot parameters for the 44~$\mu$m MMBF were first estimated from conversion factors for $\lambda_{0}$ to the cross-length and periodicity, found in Melo, et al. \cite{melo2012cross}. $\lambda_{0}$, will change in the Si medium so it must be divided by the refractive index of Si ($n_{Si}$ = 3.42) before applying the conversion factors. The estimated parameters were then optimized through the use of simulations to achieve the desired bandpass peak. These simulations are discussed in section \ref{sec:sims}. The simulation optimized cross-slot parameters are shown in Table \ref{tab:cross_params} and were used for fabrication of the 44 ~$\mu$m MMBF. 

\begin{table}[h!]
\caption{44 $\mu$m MMBF design cross-slot parameters G, K, and B (Fig. \ref{fig:mesh_geo}).}
\label{tab:cross_params}
\begin{center}       
\begin{tabular}{|l|c|} 
\hline
\rule[-1ex]{0pt}{3.5ex}  Parameter & ($\mu$m)  \\
\hline
\rule[-1ex]{0pt}{3.5ex}  G & 11.4 \\
\hline
\rule[-1ex]{0pt}{3.5ex}  K & 7.5 \\
\hline
\rule[-1ex]{0pt}{3.5ex}  B & 1.0 \\
\hline 
\end{tabular}
\end{center}
\end{table}

\section{Filter Modeling}
\label{sec:model}
\subsection{Simulations}
\label{sec:sims}
\begin{figure}[h!]
    \centering
    \includegraphics[width=.7\textwidth]{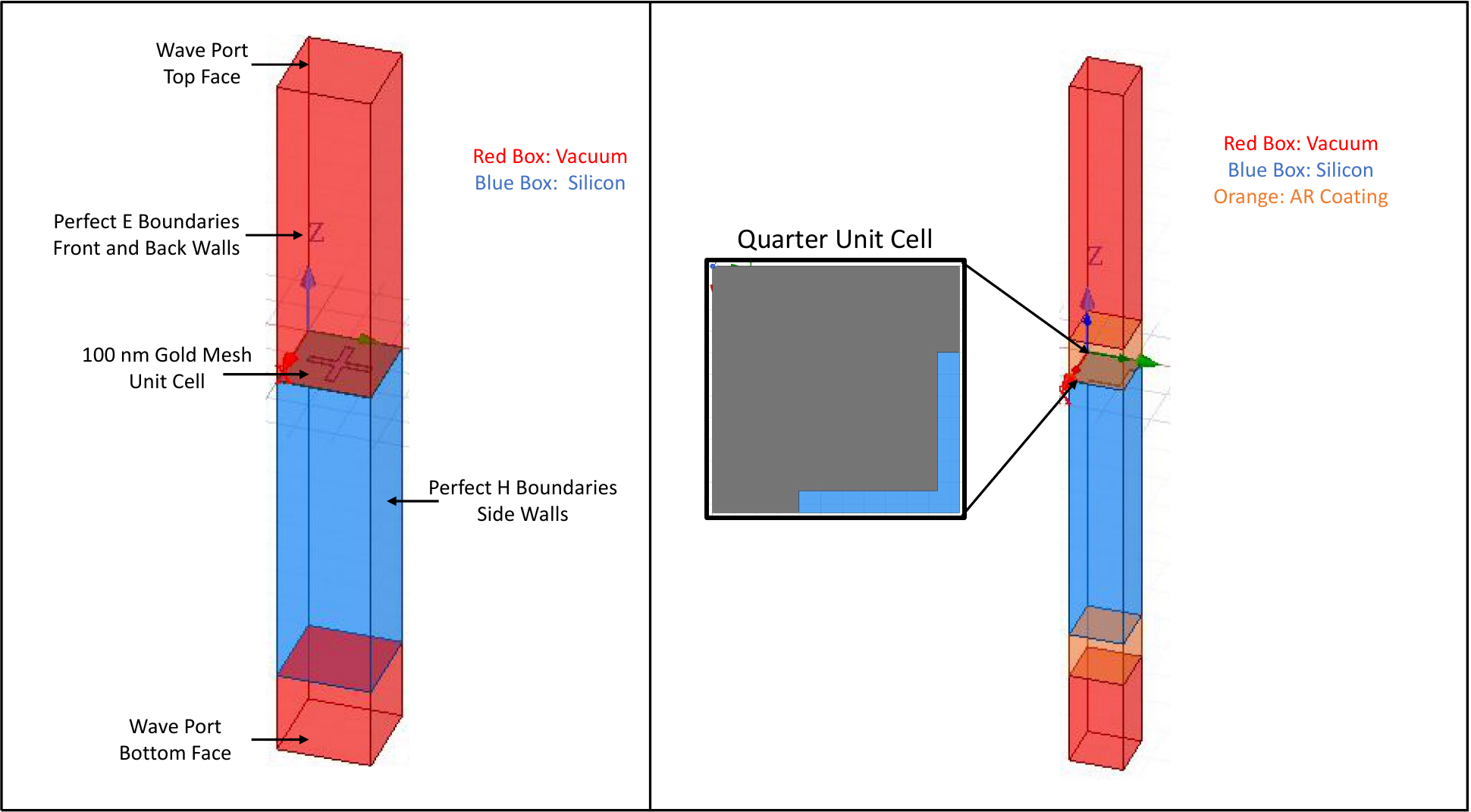}
    \caption{\textit{Left}) 3D model of an HFSS unit cell simulation for a non-AR coated MMBF. \textit{Right}) 3D model of an HFSS unit cell simulation for a double sided COC-AR coated MMBF. The quarter unit cell is used to reduce the simulation time. For illustration purposes, the vacuum, Si substrate and COC box heights are not to scale.}
    \label{fig:HFSS_struct}
\end{figure}

In order to predict and model the MMBF performance we used Ansys High Frequency Structure Simulator (HFSS)\footnote[1]{\url{https://www.ansys.com/products/electronics/ansys-hfss}} simulations. HFSS, a full-wave frequency domain electromagnetic field solver based on the finite element method, numerically solves Maxwell's equations across a specified frequency range for a specified structure geometry, material configuration, and boundary conditions. Through symmetry, an array of cross apertures in 100~nm gold on a Si substrate can be simulated by a single unit cell with perfect electric ($\vec{E}$) and magnetic ($\vec{H}$) field boundary conditions \cite{merrell2012compact, porterfield1994resonant}. The unit cell structure is shown in Fig. \ref{fig:HFSS_struct} \textit{Left}. The perfect $\vec{E}$ and $\vec{H}$ field boundary conditions are applied to the side walls of the vacuum and Si boxes. Two wave ports are used at the top of the first vacuum box and at the bottom of the second vacuum box to simulate an incident plane wave and calculate S-parameters. To reduce the simulation time, we only used a quarter of the structure shown in Fig. \ref{fig:HFSS_struct} \textit{Left}, as seen in Fig. \ref{fig:HFSS_struct} \textit{Right}. The gold was modeled as a 100~nm thick volume with a room temperature bulk conductivity of 3.33x10\textsuperscript{7} Siemens/m. The conductivity was calculated from room temperature DC resistivity measurements of the gold film, equal to 0.3~$\Omega$/sq. The Si was defined as a 537~$\mu$m thick volume with a relative dielectric permittivity, $\varepsilon_r$ = 11.7. Dielectric loss is not included in the simulations presented because we are using high-resistivity floatzone Si. The dielectric loss tangent was measured to be $\sim 1 \times 10^{-7}$, and through simulation we confirmed that this loss does not effect the transmission.

\par There is $\sim$0.7 power loss at each vacuum-Si interface. Therefore, the filters on the instruments will be at cryogenic temperatures and have anti-reflection (AR) coatings. We expect this to increase the transmission because the conductivity of gold will increase at cryogenic temperatures and vacuum-to-Si reflective losses will decrease with AR coatings. To determine the increase in transmission we ran simulations of the structure shown in Fig. \ref{fig:HFSS_struct} \textit{Right}. The cryogenic bulk conductivity is 1.33x10\textsuperscript{8} Siemens/m. This was calculated from the DC residual-resistance ratio (RRR) of the gold film, measured to be 4.0. The reflective losses at the vacuum-to-Si interfaces are reduced with $\lambda_0$/4 thick cyclic olefin copolymer\cite{wollack2020far} (COC) AR coatings. COC has a low index of refraction ($\sim$1.5) that will provide low reflective loss over a modest bandwidth at the vacuum-to-Si interfaces.  In the simulation, the COC-AR coatings are defined as $\lambda_0$/4 thick volumes with a relative dielectric permittivity of 2.37\cite{wollack2020far}. From the simulation we found that the addition of the AR coatings causes the original bandpass peak
to shift to longer wavelengths by $\sim 0.6~\mu$m because of a change in the effective capacitance of the MMBF. This shift was taken into consideration when choosing the cross-slot parameters in Table \ref{tab:cross_params}. The simulations are compared to measurements in section \ref{sec:meas_sims}. 

% \begin{figure}[h!]%
%     \centering
%     \subfloat{{\includegraphics[width=.3\textwidth]{Figs/HFFS_Struct1.pdf} }}%
%     \qquad
%     \subfloat{{\includegraphics[width=.2\textwidth, , height=6.5cm,keepaspectratio]{Figs/HFFS_Struct2.pdf} }}%
%     \subfloat{{\includegraphics[width=.3\textwidth]{Figs/HFFS_Struct4.pdf} }}%
%     \caption{}%
%     \label{fig:HFSS_struct}%
% \end{figure}

% The cross-slot parameters, which determine the transmission profile, were chosen such that when AR coatings are applied the band-pass peak will be at the desired wavelength, $\lambda_{0}$ = 44~$\mu$m. The cross-slot parameters were first estimated from conversion factors for $\lambda_{0}$ to the cross-length and periodicity, found in Melo, et al. \cite{melo2012cross}. $\lambda_{0}$, will change in the Si medium so it must be divided by the refractive index of Si, $n_{Si}$ = 3.42 before applying the conversion factors. The estimated parameters were then optimized through the use of simulations to achieve the desired band-pass peak. The simulation optimized cross-slot parameters are shown in Table \ref{tab:cross_params} and were used for fabrication of the 44 ~$\mu$m MMBF. The simulations are compared to measurements in section \ref{sec:meas_sims}. 

% \begin{table}[h!]
% \caption{Designed Cross Parameters.} 
% \label{tab:cross_params}
% \begin{center}       
% \begin{tabular}{|l|l|l|} 
% \hline
% \rule[-1ex]{0pt}{3.5ex}  Parameter & $\lambda_{0}$ = 44~$\mu$m (6.8 THz) & $\lambda_{0}$ = 24~$\mu$m (12.5 THz)\\
% \hline
% \hline
% \rule[-1ex]{0pt}{3.5ex}  g ($\mu$m) & 11.4 & 6   \\
% \hline
% \rule[-1ex]{0pt}{3.5ex}  L ($\mu$m) & 7.5 & 4.1  \\
% \hline
% \rule[-1ex]{0pt}{3.5ex}  b ($\mu$m) & 1 & 0.3  \\
% \hline 
% \end{tabular}
% \end{center}
% \end{table}

\newpage
\subsection{Transmission-Line Model}
\label{sec:tlm}
\begin{figure}[h!]
    \centering
    \includegraphics[width=.6\textwidth]{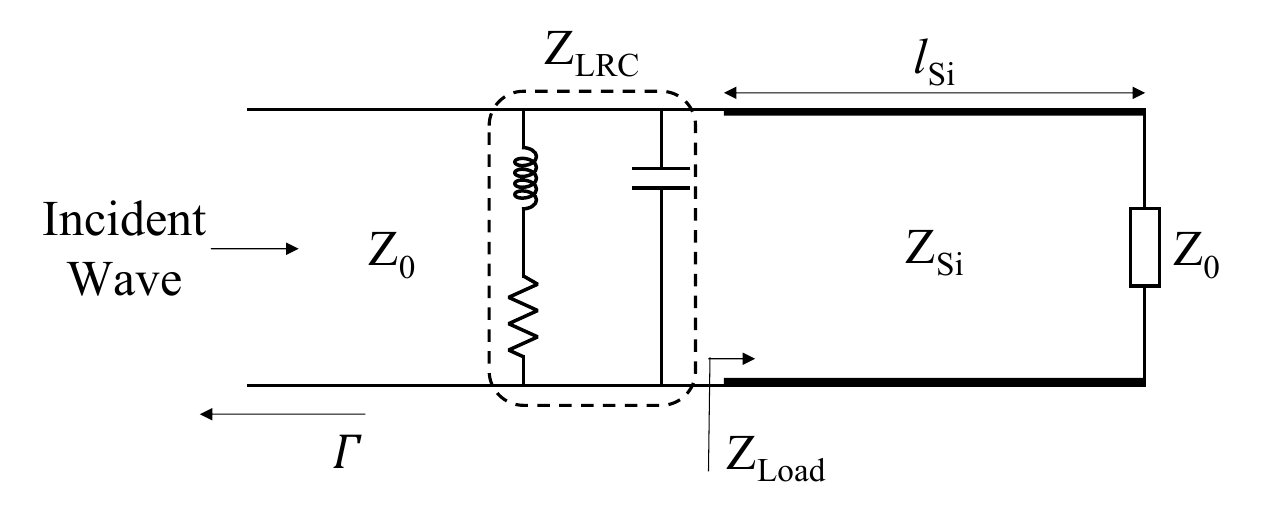}
    \caption{Transmission-line model representation of a MMBF on a Si dielectric. $Z_{0}$, is the impedance of free-space, $Z_{LRC}$ is the impedance of the MMBF, $Z_{Si}$ is the impedance of Si, $Z_{Load}$ is the load impedance at the beginning of the Si transmission line of length, $l_{Si}$. $\Gamma$, is the reflection coefficient used to calculate the fraction of incident power transmitted from the incident wave.}
    \label{fig:tlm}
\end{figure}

The metal mesh cross-slots are self-resonant structures with an impedance that can be modeled as a passive LRC circuit \cite{davis1980bandpass,porterfield1994resonant,al2019systematic}. The reactive part of the LRC circuit determines the shape of the bandpass. The real part, while still contributing to the bandwidth, mainly represents the losses in the circuit. These are due to ohmic losses and dielectric losses in the Si medium. However, since the dielectric loss is small it is not added to the current model. The transmittance of our filters can then be modeled by placing the LRC circuit in a transmission-line model, as shown in Fig. \ref{fig:tlm}. Where, $Z$\textsubscript{0} = 377 $\Omega$/sq, is the characteristic impedance of free-space. The Si is modeled as a transmission-line with length equal to thickness of the Si substrate (537 $\mu$m) with characteristic impedance, $Z$\textsubscript{Si} = $Z$\textsubscript{0}/$\sqrt{\varepsilon_r}= 110~\Omega$/sq. This model is based on the transmission-line model used by Al-Azzawi, et al.\cite{al2019systematic}. The impedance of this LRC circuit, $Z$\textsubscript{$LRC$}, is

\begin{equation}
\label{eq:zload}
Z_{LRC} = \frac{R+j\omega L}{1+j\omega RC-(\omega/\omega_0)^2}.
\end{equation}

%{\noindent \color{red} \bf Joanna:  Should there be an $\omega$ between the j and the L in the numerator?} Yes and corrected.

\noindent $R$ is the resistance, $C$ is the capacitance, $L$ is the inductance, $\omega$ is the angular frequency and $\omega_0$ is the is the circuit resonant angular frequency. In order to calculate the transmittance, the reflection coefficient, $\Gamma$, must be derived first. This is done by first calculating the load impedance, $Z_{load}$.

\begin{equation}
\label{eq:zload}
Z_{Load} = Z_{Si}\frac{Z_{0}+jZ_{Si}\tan(\beta l_{Si})}{Z_{Si}+jZ_{0}\tan(\beta l_{Si})}
\end{equation}

\noindent where $l_{Si}$ is the Si thickness (537~$\mu$m) and $\beta=\frac{2\pi}{\lambda_{Si}}$ is the wave-number. $\lambda_{Si}=\frac{\lambda}{n_{Si}}$, is the wavelength in the Si medium. Next, the equivalent impedance of the LRC circuit ($Z_{LRC}$) and load impedance must be calculated by adding them in parallel. 

\begin{equation}
\label{eq:zeq}
Z_{eq} = \frac{Z_{Load}*Z_{LRC}}{Z_{Load}+Z_{LRC}}
\end{equation}

\noindent This allows us to derive the reflection coefficient,

\begin{equation}
\label{eq:gamma}
\Gamma = \frac{Z_{eq}-Z_{0}}{Z_{eq}+Z_{0}}
\end{equation}

\noindent The transmittance, which is equal to the fraction of incident power transmitted, is evaluated following Porterfield\cite{ porterfield1994resonant} via, $T = 1-|\Gamma|^2$.
This model is fit to the HFSS simulated transmission of the MMBF to estimate $L$ and $C$ of the filter. We then compare the fit to both the simulations and the measurements in section \ref{sec:meas_tlm}.

\section{Fabrication}
\label{sec:fab}
\begin{figure}[h!]
    \centering
    \includegraphics[width=.7\textwidth]{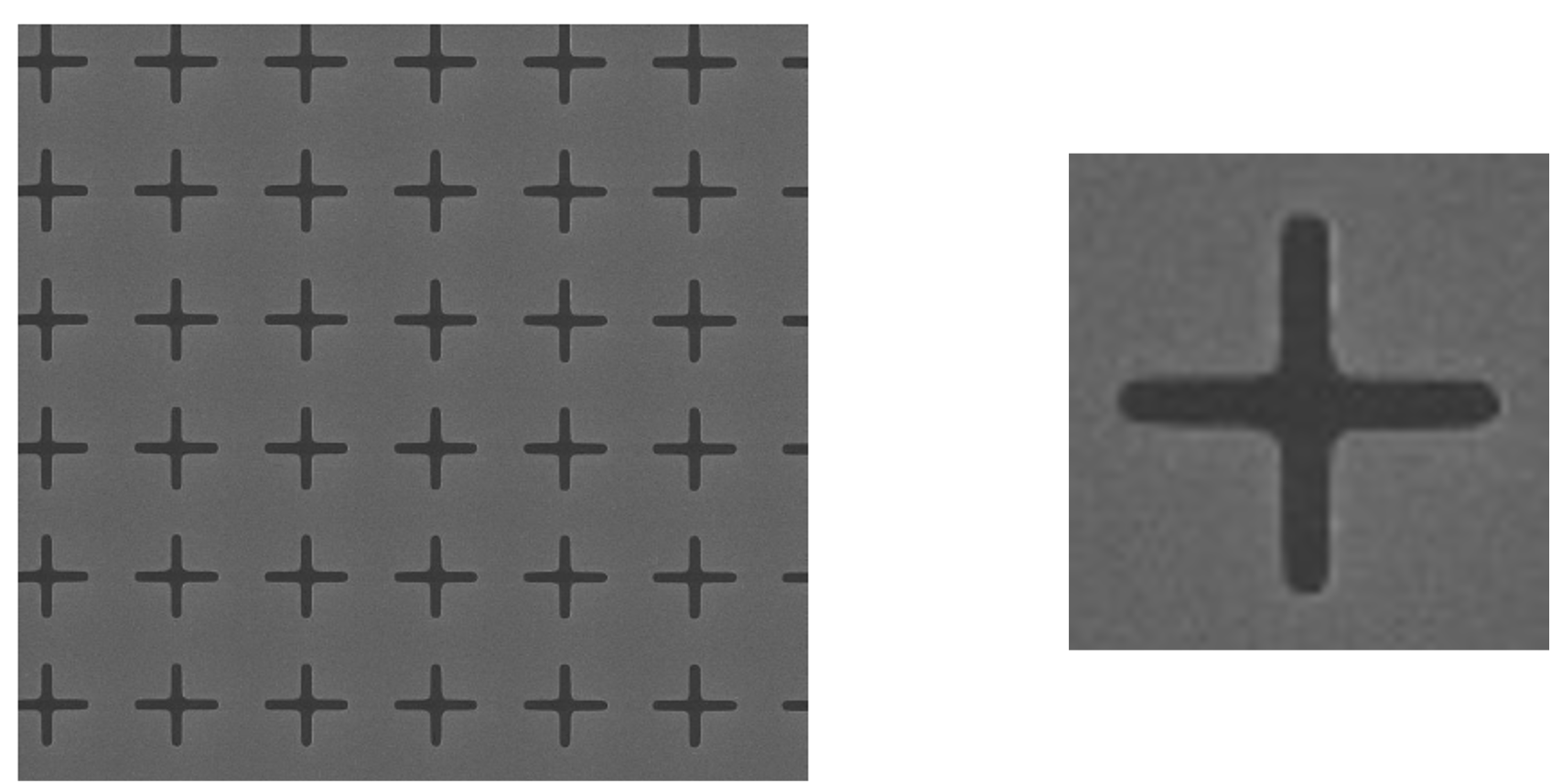}
    \caption{Scanning electron microscope (SEM) image of the fabricated 44~$\mu$m MMBF with cross-slot parameters presented in Table \ref{tab:cross_params}. The light gray is gold metal film and the dark gray is bare Si. The fabricated MMBF filter has rounded inner and outer edges with radius of curvature $\sim$0.5 $\mu$m.}
    \label{fig:fabd_filt}
\end{figure}

The 44~$\mu$m MMBF was fabricated in a simple single layer process. Double side polished intrinsic float zone silicon wafers ($\rho$ $>$ 20~k$\Omega$-cm) were coated with a 5~nm Ti adhesion layer and 100~nm thick Au layer by electron beam evaporation in the GSFC Detector Development Laboratory (DDL). The 44~$\mu$m MMBF has minimum features of 1~$\mu$m which were lithographically patterned by a Heidelburg DWL 66+ direct write laser system and a single layer of S1805 resist. The gold was etched by argon ion milling (4-Wave) and the titanium was further etched by a combination of fluorine plasma and hydrofluoric acid. Finally, the photoresist was removed by oxygen plasma and solvent cleaning. The filters were then diced into 25.4~mm by 25.4~mm squares for measurement. A scanning electron microscope image of the fabricated filter is shown in Fig. \ref{fig:fabd_filt}. The close up of one of the cross-slots shows the fabricated filter has rounded inner and outer edges, measured to have a radius of curvature of $\sim$0.5~$\mu$m. The affects of the rounded corners were investigated through simulations and are discussed in the following section.
%The LVF was patterned using lithography at NIST Gaithersburg Nanoscale Science and Technology due to the small cross aperture widths which vary from from 0.3~$\mu$m to 0.45~$\mu$m. A bilayer of bottom anti-reflection coating and i-line resist was coated, baked and exposed (ASML PAS5500-275) followed by develop. Several exposure matrix tests were done to optimize focus and intensity. We found with our original photomask designs that exposure times required to resolve the smallest features tend to over expose the wider features. A second iteration of the photomask design will account for design to experimental offsets. Exposed wafers were returned to the DDL where the lithographic AR coating was etched using an oxygen plasma to expose the gold layer. The gold was etched by argon ion milling (4-Wave) and the titanium was further etched by a combination of fluorine plasma and hydrofluoric acid. Finally, the remaining photoresist and lithographic AR coating was removed by oxygen plasma and solvent cleaning. The 44~$\mu$m MMBF has minimum features of 1~$\mu$m which allowed for the full fabrication process to be completed in the GSFC DDL with lithography exposure by a Heidelburg DWL 66+ direct write laser system and a single layer of S1805 resist without bottom lithographic AR coating. The remaining etching steps are identical to the LVF. The filters are diced into 25.4~mm by 25.4~mm squares for measurement.

\section{FTS Transmission Measurements and Modeling Discussion}
\label{sec:meas}
The transmission spectra of the 44 $\mu$m MMBF was measured using a Bruker IFS Fourier transform spectrometer (FTS) with a liquid-helium cooled bolometer. The data was collected over a spectral range from 76-650~icm (15-132~$\mu$m) with a resolution of 1~cm\textsuperscript{-1}. The measured prototype 44 $\mu$m MMBF was non-AR coated and measured at room temperature. Deployed filters will be AR-coated and operated cryogenically. We did not have measurements of AR-coated filters at cryogenic temperatures ready for these {\it Proceedings}, but we are currently working on them. Here, we present HFSS simulations and preliminary measurements of the 44~$\mu$m MMBF.

\subsection{44~$\mu$m MMBF Measurments vs Simulations}
\label{sec:meas_sims}
\begin{figure}[h!]
    \centering
    \includegraphics[width=1\textwidth]{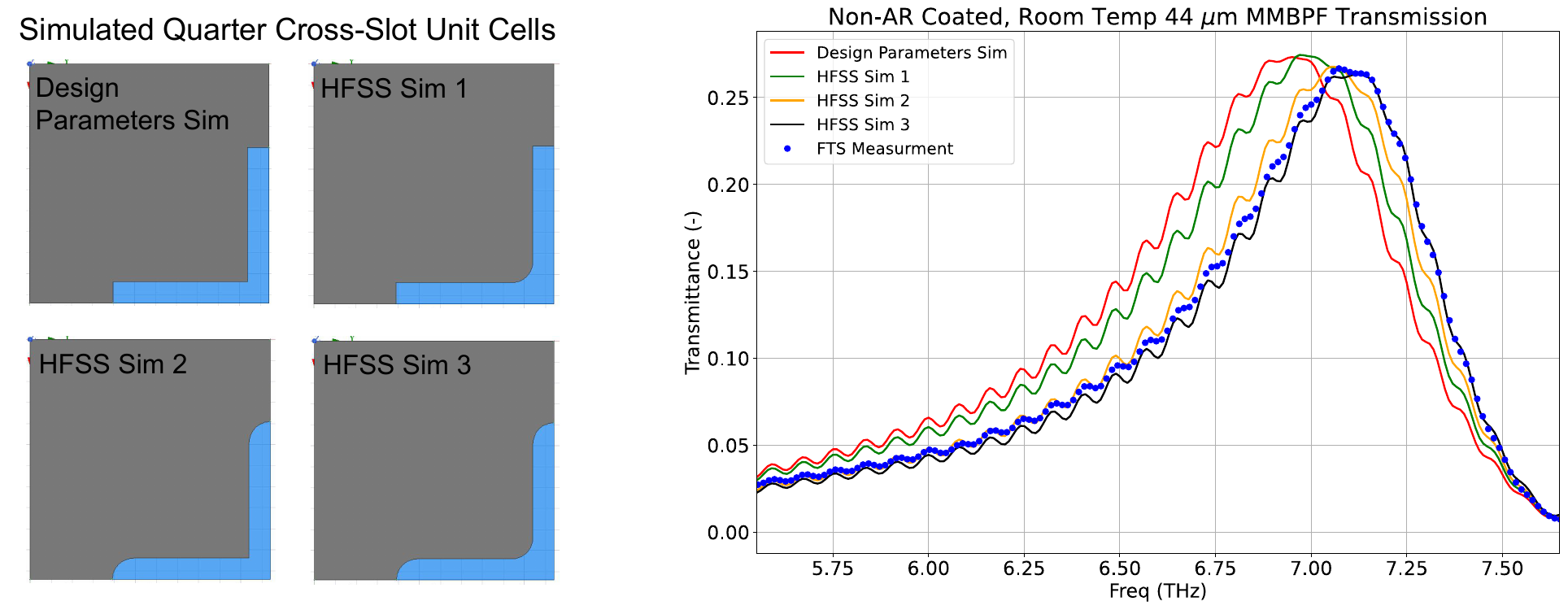}
    \caption{\textit{Left}) The HFSS simulated quarter cross-slot unit cells of the simulated transmission shown in Fig. \ref{fig:44um_meas} \textit{Right} for Design Parameters Sim, HFSS Sim 1, HFSS Sim 2, and HFSS Sim 3. \textit{Right}) HFSS simulated and FTS measured transmittance of the non-AR coated, room temperature 44~$\mu$m MMBF. Red Line: HFSS unit cell with cross-slot parameters in Table \ref{tab:cross_params}. Green Line: HFSS Sim 1 has rounded inner corners (radius of curvature = 0.5~$\mu$m) with parameters: G = 11.4~$\mu$m, B = 1~$\mu$m, K = 7.5~$\mu$m. Orange Line: HFSS Sim 2 has rounded cross-ends (radius of curvature = 0.5~$\mu$m) with parameters: G = 11.4~$\mu$m, B = 1~$\mu$m, K = 7.5~$\mu$m. Black Line: HFSS Sim 3 has rounded cross-ends and inner corners (radius of curvature = 0.5~$\mu$m) with parameters: G = 11.4~$\mu$m, B = 1~$\mu$m, K = 7.5~$\mu$m. Blue: FTS measured transmission of fabricated 44~$\mu$m MMBF. The fringing along the transmission profile is due to the Si cavity created between the metal mesh and the vacuum interface on the opposite side. The FTS resolution was small enough to barely resolve the fringes. In order for the simulations to match this, they were smoothed using a Gaussian filter to approximate the FTS apodization and resolution.}
    \label{fig:44um_meas}
\end{figure}

Fig. \ref{fig:44um_meas} shows the transmission spectra of the FTS measurement (blue dots) compared to the HFSS simulated transmission of the MMBF design cross-slot parameters (Design Parameters Sim, red solid curve). The fringing along the transmission profile is due to the Si cavity created between the metal mesh and the vacuum interface on the opposite side. The FTS resolution was small enough to barely resolve the fringes. In order for the simulations to match this, they were smoothed using a Gaussian filter to approximate the FTS apodization and resolution. The 44~$\mu$m MMBF was designed to have a band-pass peak at 7~THz with resolving power $\sim$9. However the measured filter has a bandpass peak at a slightly higher frequency, 7.1~THz, with a higher resolving power ($\sim$11) (Table \ref{tab:44um_results_1}). We determined this was due to differences in the designed MMBF's cross parameters and the fabricated cross parameters. Fig. \ref{fig:fabd_filt} shows that the fabricated 44 $\mu$m MMBF is rounded at the inner and outer edges. The inside and outside edges of the cross are rounded to about 0.5~$\mu$m or half of the cross-width. Fig. \ref{fig:44um_meas} shows the change in the simulated filter transmission when the inner edges are rounded (HFSS Sim 1, green solid curve), when the outer edges are rounded (HFSS Sim 2, yellow solid curve), and when both inner and outer edges are rounded(HFSS Sim 3, black solid line). When the the inner edges are rounded there is an increase in transmission because the area of the slot increases. There is also a slight shift in the bandpass peak to higher frequency, by 0.1~THZ. For a cross-slot with only rounded outer edges, the transmission decreases due to a decrease in the slot area with a shift to higher frequency as well but it still does not match the measured bandpass peak. The peak shifts to higher frequencies because the rounding of the edges decreases the capacitance. When both the inner and outer edges are rounded the simulated transmission aligns better with the FTS measurement. The band-pass peak is at 7.1~THz, with a transmission of 26\% and resolving power of $\sim$11. These simulations show how the filter's response is sensitive to changes in the cross-slot features and incorporating them allows us to model our measurements more accurately.
\par
\begin{figure}[h!]
    \centering
    \includegraphics[width=.6\textwidth]{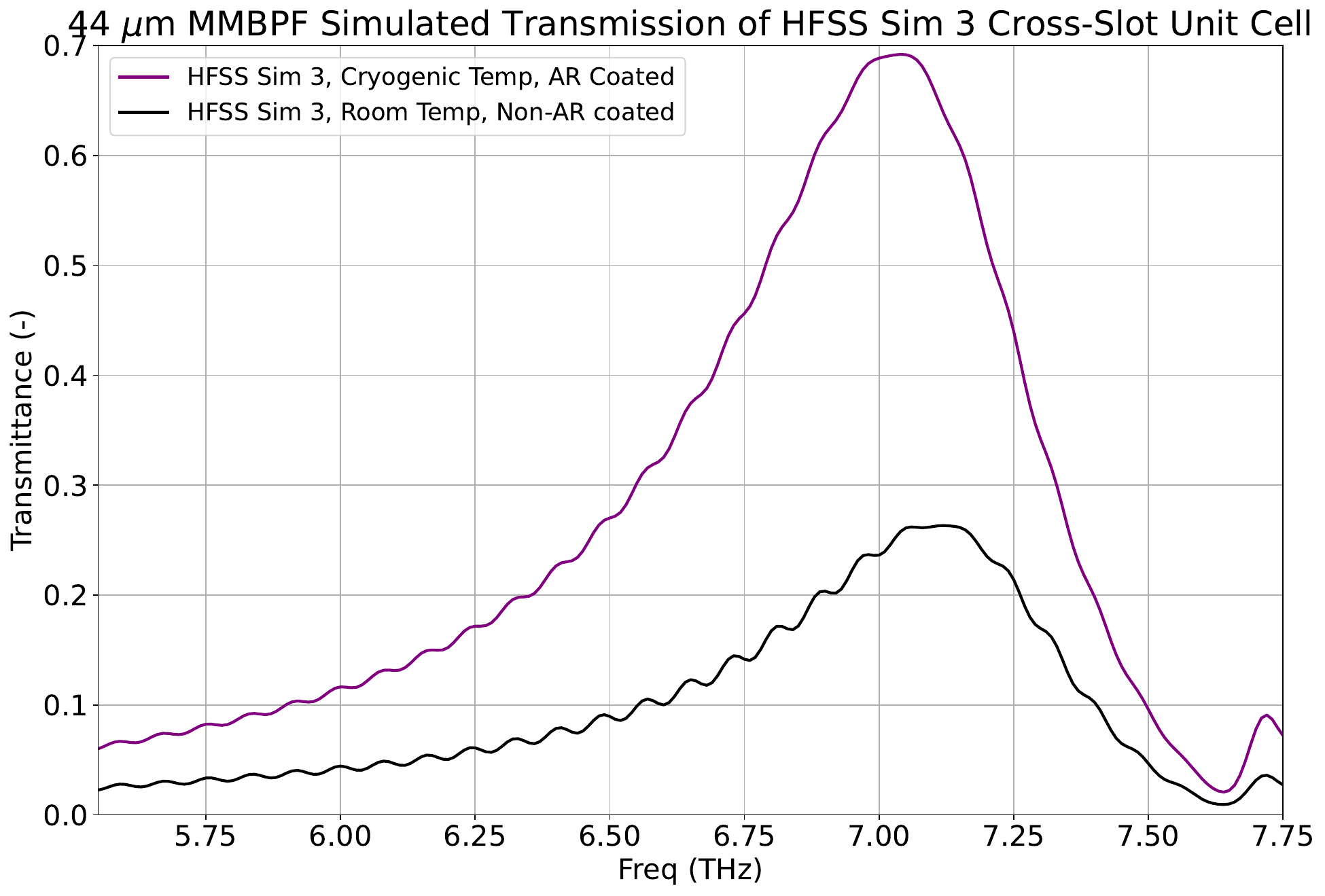}
    \caption{Purple: HFSS simulated transmission of a COC-AR coated cryogenically cooled MMBF with the cross-slot unit cell used in HFSS Sim 3. Bandpass peak = 7.0 THz (43 $\mu$m), peak transmission = 69\%, and resolving power = 10. Black: HFSS simulated transmission of non-AR coated, room temperature with the cross-slot unit cell used in HFSS Sim 3. The transmission increases by a factor of 2.63 to 69\%, due to decreased gold resistive losses at cryogenic temperatures and decreased Si reflective losses from the addition of AR coatings. The second peak at 7.72 THz is due to incident wavelengths less than the periodicity (parameter G in Table \ref{tab:cross_params}) of the the mesh grid scattered into higher order modes\cite{merrell2012compact}. The second peak occurs at the periodicity of the mesh grid (G =11.4~$\mu$m), where 7.72 THz = 39~$\mu$m/$\sqrt{\varepsilon_{r}}$=11.4~$\mu$m.}
    \label{fig:COCAR_Sim}
\end{figure}

 Fig. \ref{fig:COCAR_Sim} shows the simulated transmission of a COC-AR coated, cryogenically cooled MMBF with the cross-slot unit cell used in HFSS Sim 3. The transmission increases by a factor of 2.63 to 69\%, due to decreased gold resistive losses at cryogenic temperatures and decreased Si reflective losses from the addition of AR coatings. This is the expected transmission of our filters in an instrument.

% \begin{table}[h!]
% \caption{The band-pass peak, R, and transmission of the simulated and measured performances shown in Fig. \ref{fig:44um_meas} \textit{Right} of the 44 $\mu$m MMBF.} 
% \label{tab:44um_results}
% \begin{center}       
% \begin{tabular}{|l|l|l|l|} 
% \hline
% \rule[-1ex]{0pt}{3.5ex}  & Design Parameters HFSS Sim & Fabricated Parameters HFSS Sim & FTS Measurements\\
% \hline
% \hline
% \rule[-1ex]{0pt}{3.5ex}  Band-pass Peak (THz) & 6.9 (=43.3 $\mu$m) & 7.1 (=42.4 $\mu$m) & 7.1 (=42.4 $\mu$m)  \\
% \hline
% \rule[-1ex]{0pt}{3.5ex}  R & 9 & 10 & 10\\
% \hline
% \rule[-1ex]{0pt}{3.5ex}  Transmission (\%) & 26.7 & 26.9 & 26.7 \\
% \hline 
% \end{tabular}
% \end{center}
% \end{table}

\begin{table}[h!]
\caption{The bandpass peak, resolving power, and transmission of the simulated design parameters and FTS measured performances shown in Fig. \ref{fig:44um_meas} \textit{Left} of the 44 $\mu$m MMBF.} 
\label{tab:44um_results_1}
\begin{center}       
\begin{tabular}{|l|c|c|c|} 
\hline
\rule[-1ex]{0pt}{3.5ex} & Design Parameters HFSS Sim & FTS Measurements\\
\hline
\hline
\rule[-1ex]{0pt}{3.5ex}  Band-pass Peak (THz) & 6.9 (43 $\mu$m) & 7.1 (42 $\mu$m)\\
\hline
\rule[-1ex]{0pt}{3.5ex}  $\lambda/\Delta\lambda$ & 9 & 11\\
\hline
\rule[-1ex]{0pt}{3.5ex}  Peak Transmission (\%) & 27 & 27 \\
\hline 
\end{tabular}
\end{center}
\end{table}

% \begin{figure}[h!]%
%     \centering
%     \subfloat{{\includegraphics[width=.4\textwidth]{Figs/Sim_UnitCells.pdf} }}%
%     \qquad
%     \subfloat{{\includegraphics[width=.55\textwidth]{Figs/SimsVsMeas_44um_v4.pdf} }}%
%     \caption{\textit{Left}) The HFSS simulated quarter cross-slot unit cells of the simulated transmission shown in Fig. \ref{fig:44um_meas} \textit{Right} for Design Parameters Sim, HFSS Sim 1, HFSS Sim 2, and HFSS Sim 3. \textit{Right}) HFSS simulated and FTS measured transmittance of the non-AR coated, room temperature 44~$\mu$m MMBF. Red Line: HFSS unit cell with cross-slot parameters in Table \ref{tab:cross_params}. Green Line: HFSS Sim 1 has rounded inner corners (radius of curvature = 0.5~$\mu$m) with parameters: G = 11.4~$\mu$m, B = 1~$\mu$m, K = 7.5~$\mu$m. Orange Line: HFSS Sim 2 has rounded cross-ends (radius of curvature = 0.5~$\mu$m) with parameters: G = 11.4~$\mu$m, B = 1~$\mu$m, K = 7.5~$\mu$m. Black Line: HFSS Sim 3 has rounded cross-ends and inner corners (radius of curvature = 0.5~$\mu$m) with parameters: G = 11.4~$\mu$m, B = 1~$\mu$m, K = 7.5~$\mu$m. Blue: FTS measured transmission of fabricated 44~$\mu$m MMBF. The fringing along the transmission profile is due to the Si cavity created between the metal mesh and the vacuum interface on the opposite side. The FTS resolution was small enough to barely resolve the fringes. In order for the simulations to match this, they were smoothed using a Gaussian filter to approximate the FTS apodization and resolution.}%
%     \label{fig:44um_meas}%
% \end{figure}

\begin{table}[h!]
\caption{The band-pass peak, resolving power, and transmission of HFSS Sim 1, HFSS Sim 2, HFSS Sim3, and FTS measured performances shown in Fig. \ref{fig:44um_meas} \textit{Left} of the room temperature, non-AR coated 44~$\mu$m MMBF.} 
\label{tab:44um_results_2}
\begin{center}       
\begin{tabular}{|l|c|c|c|c|c|} 
\hline
\rule[-1ex]{0pt}{3.5ex} & HFSS Sim 1 & HFSS Sim 2 & HFSS Sim 3 & FTS Measurements\\
\hline
\hline
\rule[-1ex]{0pt}{3.5ex}  Band-pass Peak Center (THz) & 7.0 (43~$\mu$m) & 7.0 (43 $\mu$m) & 7.1 (42 $\mu$m) & 7.1 (42 $\mu$m) \\
\hline
\rule[-1ex]{0pt}{3.5ex}  $\lambda/\Delta\lambda$ & 9 & 10 & 11 & 11 \\
\hline
\rule[-1ex]{0pt}{3.5ex}  Peak Transmission (\%) & 28 & 27 & 26 & 27 \\
\hline 
\end{tabular}
\end{center}
\end{table}

\subsection{Transmission-Line Model Fitting}
\label{sec:meas_tlm}
\begin{figure}[h!]
    \centering
    \includegraphics[width=1\textwidth]{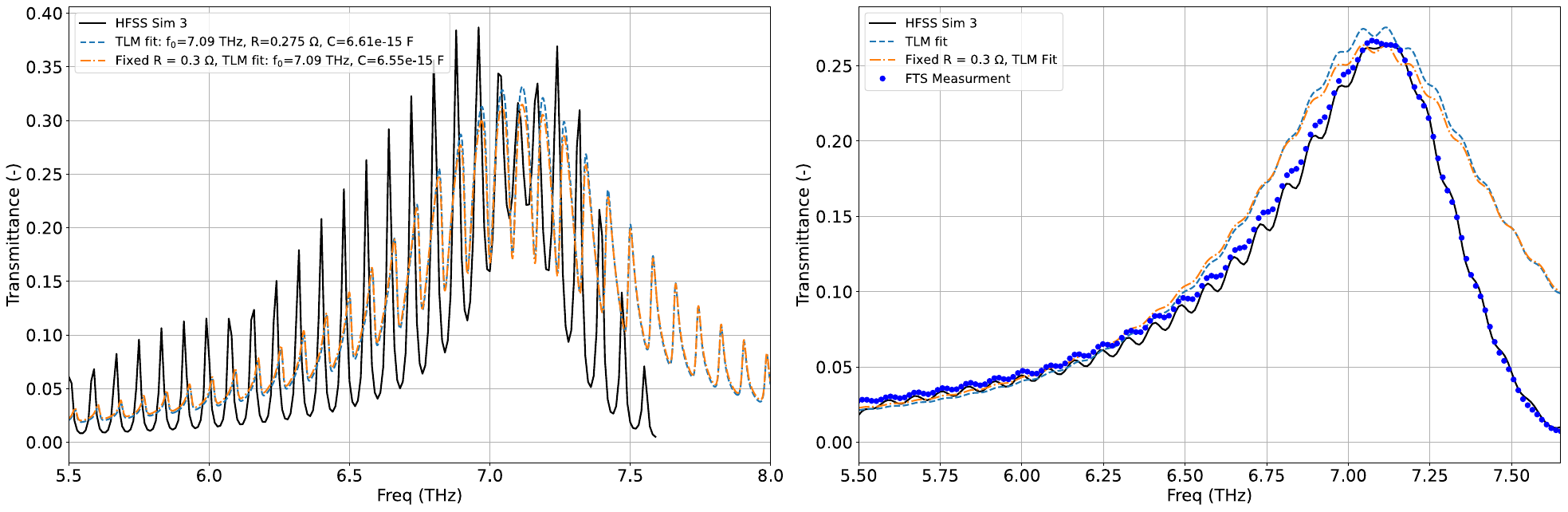}
    \caption{\textit{Left}) Transmission-line model (TLM) fits to HFSS Sim 3 in Fig. \ref{fig:44um_meas}. Black Line: Simulated transmission of HFSS Sim 3 in Fig. \ref{fig:44um_meas}. Blue Dashed Line: TLM fit to HFSS Sim 3 with three fit parameters: $f_0$, $R$ and $C$. Orange Dashed-Dotted Line: TLM fit to HFSS Sim 3 with 2 fit parameters: $f_0$ and $C$. $R$ was fixed to the measured gold DC resistance of the fabricated 44~$\mu$m MMBF, 0.3 $\Omega$/sq. The fringing along the transmission profile is due to the Si cavity created between the metal mesh and the vacuum interface on the opposite side. \textit{Right}) Smoothed TLM fits in Fig. \ref{fig:tlmfit} \textit{Left} compared to smoothed HFSS Sim 3 simulated transmission and the 44~$\mu$m MMBF measured transmission (dotted blue line). The lack of well-fit fringe amplitudes and transmission profile shown in Fig. \ref{fig:tlmfit} \textit{Left} causes the bandwidths to be larger than HFSS Sim 3 and the FTS measurement.}
    \label{fig:tlmfit}
\end{figure}

The transmission-line model was fit to HFSS Sim 3 in Fig. \ref{fig:44um_meas}. The fit has three fit parameters: $f_0$ the resonant frequency, $R$ and $C$. $L$ is not a fit parameter because it can be rewritten is terms of $C$ and $f_0$, where $L=\frac{1}{\omega_0^2 C}$. A second fit was performed with only two free parameters, $f_0$ and $C$. $R$ was fixed to match the DC resistance of the gold film, 0.3 $\Omega$/sq. The fit results are shown in Fig. \ref{fig:tlmfit} \textit{Left}. In both fits $f_0$ is fit well. However, the fringe amplitudes on either side of the bandpass peak are not well fit with our model. This means our model does not fully capture the impedance mismatch at the metal mesh and Si interface. The model is also unable to capture the asymmetry of the filter's transmission profile with the simplified LRC circuit used in Fig. \ref{fig:tlm}. Fig. \ref{fig:tlmfit} \textit{Right} compares the smoothed fits to the smoothed HFSS Sim 3 and the FTS measurement. It is clear that the lack of well-fit fringe amplitudes and transmission profile affect the bandwidths in both fits. This leads to bandwidths larger than HFSS Sim 3 and the FTS measurement. The maximum transmission is fit well when $R$ is fixed to 0.3~$\Omega$/sq. When $R$ is a free parameter it is estimated to be 2.75x10\textsuperscript{-1} $\Omega$/sq, which leads to a larger maximum transmission compared to HFSS Sim 3 and the FTS measurement. This further confirms the DC measurement of the gold film and shows that the maximum transmission is mostly dependent on ohmic losses. Table \ref{tab:TLMVals}, shows the estimated fit parameters. The estimated inductance was calculated from the resonant frequency and capacitance fits. 
Although the transmission line model does not fully capture the simulation or measurement, it is a useful tool because it enables fast estimation of transmission characteristics as a function of design parameters.
%, it helps us understand how the maximum transmission changes with resistance. It also allows us to determine how the fringe rate will change with different substrate materials and thicknesses. 
We may be able to improve the transmission line model by using a more complex circuit to represent the self-resonant metal mesh cross-slots.

% \begin{figure}[h!]%
%     \centering
%     \subfloat{{\includegraphics[width=.46\textwidth]{Figs/TLM_fit_v1.pdf} }}%
%     \qquad
%     \subfloat{{\includegraphics[width=.46\textwidth]{Figs/TLMFits_conv_v1.pdf} }}%
%     \caption{\textit{Left}) Transmission-line model (TLM) fits to HFSS Sim 3 in Fig. \ref{fig:44um_meas}. Black Line: Simulated transmission of HFSS Sim 3 in Fig. \ref{fig:44um_meas}. Blue Dashed Line: TLM fit to HFSS Sim 3 with three fit parameters: $f_0$, $R$ and $C$. Orange Dashed-Dotted Line: TLM fit to HFSS Sim 3 with 2 fit parameters: $f_0$ and $C$. $R$ was fixed to the measured gold DC resistance of the fabricated 44~$\mu$m MMBF, 0.3 $\Omega$/sq. The fringing along the transmission profile is due to the Si cavity created between the metal mesh and the vacuum interface on the opposite side. \textit{Right}) Smoothed TLM fits in Fig. \ref{fig:tlmfit} \textit{Left} compared to smoothed HFSS Sim 3 simulated transmission and the 44~$\mu$m MMBF measured transmission (dotted blue line). The lack of well-fit fringe amplitudes and transmission profile shown in Fig. \ref{fig:tlmfit} \textit{Left} causes the bandwidths to be larger than HFSS Sim 3 and the FTS measurement.}%
%     \label{fig:tlmfit}%
% \end{figure}

\vspace{\baselineskip}

\begin{table}[h!]
\caption{TLM fit parameter values for $f_0$, $R$, $C$ and $L$. The middle column contains the fit parameters from the blue dashed line in Fig. \ref{fig:tlmfit} \textit{Left}. The right column contains the fit parameters from the orange dashed-dotted line in Fig. \ref{fig:tlmfit} \textit{Left}, where R was fixed.} 
\label{tab:TLMVals}
\begin{center}       
\begin{tabular}{|l|c|c|} 
\hline
\rule[-1ex]{0pt}{3.5ex} & TLM Fit & Fixed R, TLM Fit\\
\hline
\hline
\rule[-1ex]{0pt}{3.5ex}  $f_0$ (THz) & 7.09 & 7.09\\
\hline
\rule[-1ex]{0pt}{3.5ex}  $R$ ($\Omega$/sq) & 0.275 & 0.300\\
\hline
\rule[-1ex]{0pt}{3.5ex}  $C$ (fF) & 6.61 & 6.55 \\
\hline
\rule[-1ex]{0pt}{3.5ex}  $L$ (fH) & 76.2 & 76.9\\
\hline
\end{tabular}
\end{center}
\end{table}

\section{Conclusion and Future Work}
\label{sec:conc}
We have successfully fabricated, measured, and modeled a non-AR coated, room temperature 44~$\mu$m MMBF. The transmission at room temperature and non-AR coated was measured to be 27\% with a resolving power of 11. When COC-AR coated on both sides the transmission is expected to increase to 69\% with a resolving power of 10. The transmission-line model we developed did not fit the simulations well. However, it does helps us understand how the maximum transmission changes with resistance and determine how the fringe rate changes with substrate material and thicknesses. We will continue to improve this transmission line model by modifying the circuit used to model the cross-slot resonance. 
\par Our next steps are to characterize cryogenically cooled ($\sim$5 K) COC-AR coated 44~$\mu$m filters, LVFs and DVFs. With the simulation methods described and a similar fabrication process, we have begun work to develop LVFs and DVFs for the BEGINS instrument from 25-65~$\mu$m and the NFR. Like the 44 $\mu$m MMBF, they are made of 100~nm thick gold with cross-slots on a Si substrate. The LVFs and DVFs will have a resolving power of $\sim$7. It should be possible to span the whole wavelength range by scaling the cross-slot parameters, while still preserving the resolving power\cite{porterfield1994resonant,merrell2012compact}.

% \section{Acknowledgments}
\acknowledgments % equivalent to \section*{ACKNOWLEDGMENTS} 
% Research performed in part at the NIST Center for Nanoscale Science and Technology
This work is supported by: Future Investigators in NASA Earth and Space Science and Technology (FINESST) grant to Joanna Perido, NASA Astrophysics Research and Analysis (APRA) grant to the California Institute of Technology, JPL, and the University of Colorado-Boulder
and a NASA Goddard Space Flight Center Internal Research and Development (IRAD) grant.
NFC was supported by an appointment to the NASA Postdoctoral Program at Goddard Space Flight Center, administered by the Oak Ridge Associated Universities under contract with NASA.

% \section{References}

% Note: If compiling with LaTeX+dvipdf, please ensure images generated from 
% other software packages have their bounding boxes set correctly.

% References
\bibliography{report} % bibliography data in report.bib
\bibliographystyle{spiebib} % makes bibtex use spiebib.bst

\end{document}